\documentclass[10pt,preprint]{aastex}

\shorttitle{The M31 Giant Southern Stream}
\shortauthors{Kalirai et al.}

\begin{document}


\title{Kinematics and Metallicity of M31 Red Giants: The Giant Southern 
Stream and Discovery of a Second Cold Component at R = 20 kpc}


\author{Jasonjot S.\ Kalirai\altaffilmark{3,4},
Puragra Guhathakurta\altaffilmark{3},
Karoline M.\ Gilbert\altaffilmark{3},
David B.\ Reitzel\altaffilmark{5}, \\
Steven R.\ Majewski\altaffilmark{6},
R.\ Michael Rich\altaffilmark{5},
and
Michael C.\ Cooper\altaffilmark{7}}


\altaffiltext{1}{Data presented herein were obtained at the W.\ M.\ Keck
Observatory, which is operated as a scientific partnership among the
California Institute of Technology, the University of California, and the
National Aeronautics and Space Administration.  The Observatory was made
possible by the generous financial support of the W.\ M.\ Keck Foundation.}
\altaffiltext{2}{Based on observations obtained with MegaPrime/MegaCam, a
joint project of CFHT and CEA/DAPNIA, at the Canada-France-Hawaii Telescope
(CFHT) which is operated by the National Research Council (NRC) of Canada,
the Institut National des Science de l'Univers of the Centre National de la
Recherche Scientifique (CNRS) of France, and the University of Hawaii.}
\altaffiltext{3}{University of California Observatories/Lick Observatory, 
University of California at Santa Cruz, 1156 High Street, Santa Cruz, 
California 95064, USA; jkalirai@ucolick.org}
\altaffiltext{4}{Hubble Fellow}
\altaffiltext{5}{Department of Astronomy, University of California at Los Angeles, 
Box 951547, Knudsen Hall, Los Angeles, California 90095, USA}
\altaffiltext{6}{Department of Astronomy, University of Virginia, P.\ O.\ Box
3818, Charlottesville, Virginia 22903, USA} 
\altaffiltext{7}{Astronomy Department, 601 Campbell Hall, University of
California at Berkeley, Berkeley, California 94720, USA}


\begin{abstract}

We present spectroscopic observations of red giant branch (RGB) stars in the
Andromeda spiral galaxy (M31), acquired with the DEIMOS instrument on the
Keck~II 10-m telescope.  The three fields targeted in this study are in
the M31 spheroid, outer disk, and giant southern stream.  In this paper, we focus
on the kinematics and chemical composition of RGB stars in the stream field
located at a projected distance of $R=20$~kpc from M31's center.  A mix of
stellar populations is found in this field.  M31 RGB stars are isolated from
Milky Way dwarf star contaminants using a variety of spectral and photometric
diagnostics.  The radial velocity distribution of RGB stars displays a clear
bimodality --- a primary peak centered at $\bar{v_1}=-513$~km~s$^{-1}$ and a
secondary one at $\bar{v_2}=-417$~km~s$^{-1}$ --- along with an underlying
broad component that is presumably representative of the smooth spheroid of
M31.  Both peaks are found to be dynamically cold with intrinsic velocity
dispersions of $\sigma(v)\approx16$~km~s$^{-1}$.
The mean metallicity and metallicity dispersion of stars in the two peaks is
also found to be similar: $\rm\langle[Fe/H]\rangle \sim -0.45$ and
$\rm\sigma([Fe/H])=0.2$.  The observed velocity of the primary peak is
consistent with that predicted by dynamical models for the stream, but there
is no obvious explanation for the secondary peak.  The nature of the secondary 
cold population is unclear: it may represent: (1) tidal debris from a satellite 
merger event that is superimposed on, but unrelated to, the giant southern stream; 
(2) a wrapped around component of the giant southern stream; (3) a warp or overdensity 
in M31's disk at $R_{\rm disk}>50$~kpc (this component is well above the outward 
extrapolation of the smooth exponential disk brightness profile).

\end{abstract}





\section{Introduction} \label{intro}

Non-baryonic matter is now known to comprise a significant fraction (23\%) of
the total matter/energy content of the Universe (Spergel et~al.\ 2003).  It
is generally believed that this matter is most likely in the form of weakly
interacting cold dark matter (CDM).  Cosmological simulations suggest that
the large-scale distribution of matter in the Universe is consistent with CDM
predictions (i.e., observations of galaxy clustering).  Simulations also
suggest that the evolution of this dark matter imprints a signature on much
smaller scales.  In fact, it is expected that CDM will cluster
gravitationally on subgalactic scales.  Within the framework of hierarchical
formation scenarios (Searle \& Zinn 1978), it is the merging and
accretion of these subhalos that is thought to build up massive galaxies such
as the Milky Way and our neighbor the Andromeda spiral galaxy (M31).

The cores of the CDM subhalos that fall into massive potentials to form large
galaxies are expected to remain intact, even after experiencing strong tides
through several orbital timescales around the host galaxy (Hayashi et~al.\
2003).  A fraction of the subhalos will have also likely experienced star
formation either prior to, or during assimilation with the host.  A parent
galaxy the size of the Milky Way or M31 may be expected to have cannibalised,
or at the least be host to, 100--500 dwarf galaxy mass systems over its
lifetime (e.g., Klypin et~al.\ 1999; Moore et~al.\ 1999; Bode, Ostriker, \& 
Turok\ 2001).

Numerical simulations suggest that the accretion of dwarf galaxies by massive
hosts should leave fossil signatures in the form of tidal streams around
massive galaxies (Johnston 1998; Bullock, Kravtsov, \& Weinberg 2001).  The 
Sagittarius dwarf galaxy (Ibata, Gilmore, \& Irwin 1994; Majewski et al. 2003) is 
an excellent example of such a merger in our own Galaxy.  Other merger events
are also seen in the Milky Way, such as the Magellanic stream (Mathewson,
Cleary, \& Murray 1974) and the Monoceros stream (Yanny et~al.\ 2003).  The
recent discovery of the giant stellar stream to the south of M31 (Ibata
et~al.\ 2001) confirms that tidal streams are in fact a remnant of the 
accretion process, and that this process is still occurring 
today in galaxies.  Studying the dynamics, composition, and structure of
these streams will help produce a clear picture of the recent accretion
history of these host galaxies and further help constrain what fraction of halos 
may originate in these accretion events.  In some ways such studies are more 
easily carried out in M31 than the Milky Way since our location within the latter 
cause local streams to have a larger angular extent and lower stellar density 
on the sky.  The giant stellar stream of M31 is of particular interest as it is 
easily seen in optical star-count maps spanning over a very large radial extent 
($>$125 kpc), and therefore can set important constraints on the potential of M31 
(e.g., Geehan et~al.\ 2005)

\section{Previous Studies of the M31 Giant Southern Stream} \label{previous}

Since its discovery by Ibata et~al.\ (2001), the giant southern stream in M31
has been the focus of many scientific papers.  The brightest red giant branch
(RGB) stars in the stream have $I_0\approx20.5$, and therefore have been
targeted with both imagers and spectrographs in ground- and space-based studies.  The spatial 
extent of the stream has now been mapped in the south-east direction from an outermost field 
at a projected radial distance of $~\sim$60 kpc to an inner field at $\sim$10 kpc from the center 
of M31 (Ferguson et~al.\ 2002; Ibata et~al.\ 2004).  Based on fits to the tip of the RGB, the 
line of sight distances to stars in the outer regions of the stream indicate 
that that these regions are located $\sim$100 kpc behind M31 (McConnachie et~al.\ 2003).  
On the north-west side, the stream can be traced to a point where it is $\sim$40 
kpc in front of M31 (McConnachie et~al.\ 2003).

Spectroscopic studies along the path of the stream have provided both
abundance and kinematical measurements for its constituent stars as a function
of radial distance from M31's center.  Ibata et~al.\ (2004) obtained
Keck/DEIMOS spectra in four fields spanning a range of radial distances in
the southern quadrant of M31.  Their data show that at the outer regions 
of the stream, stars are moving with a heliocentric velocity of $v=-320$~km~s$^{-1}$ 
(i.e., at approximately the systemic velocity of M31).  However, closer to
M31 (in Ibata et~al.'s field~6 at a projected radius of $R\sim25$~kpc), the stream stars 
are blueshifted by an additional 160~km~s$^{-1}$.  In their innermost field ($R\sim12$~kpc), 
Ibata et~al.\ only find 2--4 likely stream stars moving with a heliocentric velocity of 
$v_{\rm hel}$ = $\rm-$520 km~s$^{-1}$.  Over its radial extent, the velocity distribution of 
stream stars shows a very sharp negative edge based on these studies. The velocity dispersion 
is found to be very narrow, $\sigma_v=11\pm3$~km~s$^{-1}$.  Guhathakurta et~al.\ (2005a) 
provide further Keck/DEIMOS observations of 45 bona-fide stream members located in a 
field at a projected distance of $\sim$33 kpc from the center of M31 (between two of the 
Ibata et~al.\ 2004 pointings).  By deconvolving the general halo population from the stream, 
they find that the stream is both more metal rich ($\langle$[Fe/H]$\rangle$ = $\rm-$0.51) 
and contains a much smaller metallicity dispersion (0.25 dex) than the general halo.  At this 
location, the heliocentric velocity of the stream is found to be $v$ = $\rm-$458 km~s$^{-1}$ with a 
low internal line-of-sight velocity dispersion ($\sigma_v=15^{+8}_{-15}$~km~s$^{-1}$).

The Andromeda giant stream has now been studied with imaging/spectroscopy over a 125 kpc 
radial extent.  This work has shed light on the dynamics, orbit, and metallicity of 
the stream, and put constraints on the progenitor galaxy (Fardal et~al.\ 2005).  
From analyzing the width and velocity dispersion of the debris, Font et~al.\ (2005) find that 
the progenitor of the stream must have been a dwarf galaxy with a mass of at least 
10$^{8}$ $M_\odot$.  The orbit of this progenitor is believed to have been 
highly eccentric and close to edge on, passing very close to the nucleus of M31 
(within 2~kpc of M31's center --- Ibata et~al.\ 2004; Font et~al.\ 2005; Fardal
et~al.\ 2005).  The high surface brightness of the 
stream indicates that this passage occured within the past few Gyr at most
(Ibata et~al.\ 2004; Fardal et~al.\ 2005).  

Despite the wealth of knowledge that we have on the Andromeda stream, many important 
questions remain unanswered.  If it still exists, the progenitor galaxy has yet to 
be identified although there are several interesting candidates.  These include the claimed 
tidally distorted satellite galaxy And VIII (Morrison et~al.\ 2003) or the concentration found 
in the distribution of M31 planetary nebulae (Merrett et~al.\ 2003).  Also unclear is whether the 
stream does in fact extend 
to the north-west side of M31, in front of the galaxy, as suggested by McConnachie 
et~al.\ (2003).  No spectroscopic observations have yet been obtained in this region of 
M31.  Finally, given their very small pericenter, stream stars have almost certainly been, 
and will continue to be, stripped from their progenitor and subsequently injected into 
the general halo of M31 (Ibata et~al.\ 2004).  This ``pollution'' potentially makes it very 
difficult to interpret age/metallicity measurements in the ``smooth'' halo and needs to 
be characterized.  For example, the recent discovery of an intermediate age population 
of stars in the spheroid of M31 (Brown et~al.\ 2003) may represent stream contamination 
(Ferguson et~al.\ 2005).  

To answer some of these important questions, we have begun a large program to
obtain Keck/DEIMOS spectroscopy of M31 RGB stars.  The three spectroscopic
fields presented in this paper were chosen to directly overlap recent
ultra-deep {\it Hubble Space Telescope\/} ({\it HST})/Advanced Camera for
Surveys imaging pointings of M31 (P.I. T. Brown - Cycle 13):
(1) a spheroid field on the south-east minor axis located at $R=12.2$~kpc (H11),
(2) a stream field located at $R=21.3$~kpc (H13s), and (3) a disk field on the north-east major 
axis at $R=25.8$~kpc (H13d).  The distances given here apply to our spectroscopic field centers and 
not the {\it HST}/ACS field centers.  In this contribution, we present the observations and 
data reduction for all three fields (\S\,\ref{data}), as well as a brief discussion on 
obtaining a clean sample of bonafide M31 RGB stars in each field (\S\,\ref{cleansample}).  The 
imaging observations are analyzed in \S\,\ref{imageanal}, where we present both 
color-magnitude diagrams (CMDs) and derive photometric metallicities for our sample.  The 
spectroscopic observations are presented in \S\,\ref{spectraanal}.  This includes 
determining the chemical composition and kinematics of stars in each field, 
with particular focus on the stream observations.  These data are analyzed and discussed 
in \S\,\ref{discdisc}.  Ongoing and future projects are highlighted in
\S\,\ref{future} and the conclusions of the present study are given in
\S\,\ref{conclusion}.  Forthcoming papers will present additional analysis of the 
data in fields H11 and H13d.

\section{Observations and Data Reduction}\label{data}

\begin{figure}
\begin{center}
\resizebox{10cm}{!}{\rotatebox{270}{\includegraphics{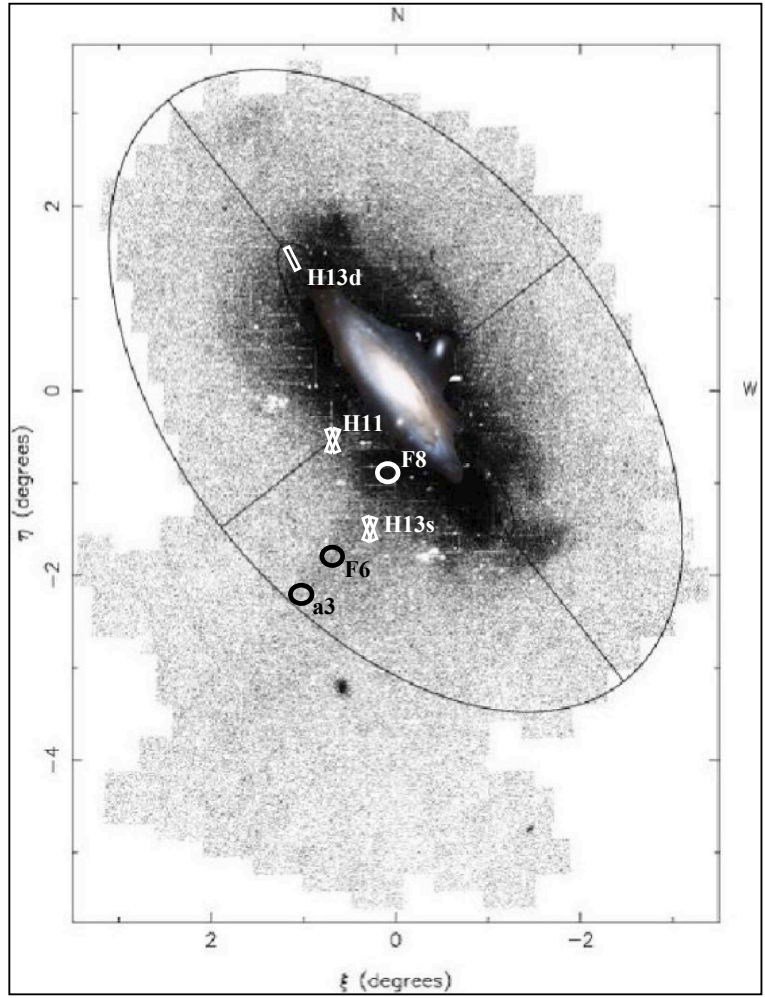}}}
\end{center}
\caption{The location of our three spectroscopic fields H13s, H11, and H13d, are shown 
as rectangles superimposed on a star-count map of M31 (Ibata et~al.\ 2005).  Approximate 
locations of other studies of the giant southern stream are also shown as circles (to be discussed 
later).  A scaled DSS image of M31 is shown at the centre of the star-count map.}
\label{fig:fields}
\end{figure}

The locations of the stream (H13s), spheroid (H11), and disk (H13d) spectroscopic 
fields presented in this study are shown as white rectangles superimposed on a
star-count map of M31 (Ferguson et~al.\ 2002; Ibata et~al.\ 2005) in
Figure~\ref{fig:fields}.  For H13s and H11, two spectroscopic observations were 
obtained at the same right ascension ($\alpha$) and declination ($\delta$), but with
different mask position angles.  For H13d, the same position angle was used for both masks.  
The giant southern stream can be clearly seen extending radially outward in the south-east 
region.  Fields studied in other papers as discussed earlier are shown as open circles 
(see \S \ref{interpdisk}).

\subsection{Imaging Fields and Astrometry}\label{preimaging}

RGB candidates were selected as targets for the Keck/DEIMOS spectrograph from
CFHT/MegaCam photometry for the H13s and H11 fields, and from Keck/DEIMOS
imaging for the H13d field.  The MegaCam pointings, each covering
1~degree$^2$, are at relatively small radial distances from M31: one is
directly along the south-east minor axis and the other is centered on the
giant southern stream.  The MegaCam images overlap with the much smaller 
Keck/DEIMOS fields ($16'\times4'$).  For the wide-field photometry, we 
obtained three 1160\,s exposures with MegaCam in each of the $g'$ and $i'$
filters, for both pointings.  All MegaCam images were obtained in photometric
conditions, with sub-arcsecond seeing (most images have
$0\farcs6$--$0\farcs8$) and airmasses less than 1.3. 
The $g'$ and $i'$ photometry from CFHT/MegaCam was converted in Johnson-Cousins $V$ and $I$ using standard star fields.

Astrometric solutions were built from the CFHT images using several steps 
(P.\ Stetson 2004, private communication).  First, the US Naval Observatory 
USNO-A2.0 Guide Star Catalog (Monet et~al.\ 1998) was used to extract all 
standards falling within either of the MegaCam pointings.  Stellar positions 
were then measured on overlapping Digitized Sky Survey plates using a modernized 
version of the Stetson (1979) software.  Coordinates in these files, and those 
from an ALLFRAME analysis of the CFHT/MegaCam images, were all put onto a common 
reference frame using ten-parameter cubic fits in $x$ and $y$ within DAOMASTER.  The 
resulting files contain $x$ and $y$ in arcseconds, increasing to the east and north,
respectively, based on a primary reference frame of the USNO-A2.0 positions.  The 
accuracy of the coordinates from this transformation, based on fitting the residuals, 
is better than 0$\farcs$1 (see \S\,2.2 of Stetson, McClure, \& VandenBerg 
2004 for more information). 

Relevant sections of the 1~degree$^2$ CFHT image that overlapped the
Keck/DEIMOS masks were extracted from each of the three 1160\,s 
exposures, for each filter.  These were bias subtracted, flat fielded,
shifted and combined using the FITS Large Images Processing Software (FLIPS)
developed at CFHT (J.-C.\ Cuillandre, 2001, private communication).  More
information on FLIPS can be found in \S\,3.1 of Kalirai et al.\ (2001).  For
the processed $g'$ and $i'$ images, we used SExtractor (Bertin \& Arnouts
1996) to produce a catalog of all stars, and applied the above
transformation to derive $\alpha$ and $\delta$ (J2000) using the standard
{\sc iraf}\footnote{{\sc iraf} is distributed by the National Optical 
Astronomy Observatories, which are operated by the Association of
Universities for Research in Astronomy, Inc., under cooperative agreement
with the National Science Foundation.} \normalfont tasks {\sc CCMAP} and {\sc
CCTRAN}\normalfont.

For H13d, 1--300\,s imaging exposure was obtained using Keck/DEIMOS in July 
2004 by replacing the grating with a mirror.  This field was also obtained at 
low airmass (1.11) and good seeing (0$\farcs$7).  An astrometric solution was built 
for each of the 4 CCDs based on manually aligning 10--15~USNO standard stars per CCD 
with their observed centers.  The derived transformation was applied to the remaining
stars, giving $\alpha$ and $\delta$ (J2000).

\subsection{Multi-object Spectroscopic Masks}\label{masks}

Target selection and slitmask design for Keck/DEIMOS spectroscopy were
carried out using a very similar method to that described in Guhathakurta
et~al.\ (2005a).  Summarizing, we first assign a priority to each object
based primarily on its magnitude ($20.5<I_0<22.5$) and stellarity (morphology
index from SExtractor used to distinguish stars from galaxies).  These
priorities are fed into the {\bf
dsimulator}\footnote{\url{http://www.ucolick.org/$\sim$phillips/deimos\_ref/masks.html}} 
software program, which builds the most efficient mask by maximizing the
number of top priority objects first, and then subsequently adding lower 
priority stars.  These lower priority stars are primarily those with fainter 
magnitudes, and will therefore have a lower signal-to-noise (S/N) spectra.  
The program prevents any overlapping slits and ensures both 
a small gap 0$\farcs$5 in the spatial direction between any two slits and a
minimum length of 6$''$ for each slitlet.  Predicted locations of the
inter-CCD gaps and vignetted regions are avoided.  All slits were produced
with a width of 1$''$.  This procedure allowed for $\sim$150 slits to be cut
on each mask.  We note that we have yet to recover spectra for $\sim12.5\%$
of the slits in our H13s and H11 fields due to problems with slit extraction.  
The missing slits are all located on two of the DEIMOS eight CCDs and it is 
yet unclear why the extractions for these objects failed.

Two spectroscopic masks were constructed for each of H13s, H11, and H13d.  For 
H13s and H11, we offset the position angle by $\pm$21.0 degrees for the two 
pointings.  The central $\alpha$ and $\delta$ are identical, thereby allowing 
for a small overlap in the database of available stars for each mask.  Two stars 
in each of these fields were observed twice.  For H13d, a position angle of $\rm+$27.8 was 
used for both masks to allow two spectra to be obtained for 53 objects.

\subsection{Spectroscopic Observations}\label{specobs}

The six~multislit masks (two for each of the H11, H13s, and H13d fields) were
observed using the Keck~II 10-m telescope and DEIMOS spectrograph on 2004
September 20 (UT).  We obtained $3\times20$~min exposures for each mask at
relatively low airmass ($<1.3$).  The grating used for the spectroscopy has
1200~lines mm$^{-1}$ and a dispersion of $\rm0.33\AA$~pixel$^{-1}$.  The
grating tilt was chosen such that the central wavelength was about
$\rm7800\AA$, thereby providing spectral coverage in the range
$\approx6400$--$\rm9100\AA$.  The resolution of the spectra is $\rm1.3\AA$
(FWHM) for typical 0$\farcs$8 FWHM seeing.

The targeted fields are shown in Figure~\ref{fig:fields} superimposed on 
a star-count map of M31.  For the stream field, H13s, we observe the two pointings 
at a radial distance of 21.3 kpc from the center of M31.  For the spheroid field, H11, 
we observe both pointings on the south-east minor axis located at $R$ = 12.2 kpc.  For 
the disk field, H13d, we observe both pointings on the north-east major axis at $R$ = 25.8 
kpc.

Table~1 summarizes the spectroscopic observations, positions of the fields, and 
numbers of targets observed on each mask.

\subsection{Spectroscopic Data Reduction}\label{specdatared}

The six DEIMOS masks were processed using the {\bf 
spec2d}\footnote{\url{http://astron.berkeley.edu/$\sim$cooper/deep/spec2d/primer.html}} 
software pipeline (v1.1.4) developed by the DEEP2 team at the University of 
California-Berkeley (UCB).  A summary of the steps involved in the data 
reduction including the flat-fielding, fringe correction, wavelength calibration, 
sky subtraction, cosmic-ray rejection, and extraction can be found in 
Guhathakurta et~al.\ (2005a).

\subsection{Velocity Measurements and Quality Assessment}\label{velocityquality}

The final extracted spectra are next processed through the {\bf 
spec1d}\footnote{\url{http://astron.berkeley.edu/$\sim$cooper/deep/spec1d/primer.html}} 
software pipeline, also developed by the DEEP2 team at UCB.  This software 
cross-correlates the observed spectrum with a large database of both template 
stellar spectra (multiple A, B, F, G, K, L, M type, and Carbon star) and template emission- and 
absorption-line galaxy spectra to determine the red/blueshift for all objects.  The 
results are displayed in the visual 
inspection software {\bf zspec} (developed by D.\ Madgwick for the DEEP2 survey).  In 
rare cases, the automated cross-correlation failed despite the presence of obvious 
spectral features upon visual inspection (e.g., due to artificats from poorly 
subtracted night sky lines).  For these cases (7\% of all targets with such features), the 
radial velocity was manually determined using any of several spectral features: 
the Ca\,{\sc ii} triplet, the Na\,{\sc i} doublet (dwarfs), and/or the $\rm7100\AA$ TiO 
band.  The typical signal-to-noise of these spectra is $\sim$10 per pixel.  The 
velocity uncertainty from the cross-correlation is empirically estimated to be $\sim$10 
km~s$^{-1}$ from the 57 objects that were observed twice (see \S\,2.4 of Gilbert 
et~al.\ 2005, in preparation) in H13s, H11, and H13d.

A heliocentric velocity correction was applied to each of the fields using the 
{\sc iraf} task RVCOR.  The corrections were determined to be +12.17~km~s$^{-1}$ for 
H13s, +12.45~km~s$^{-1}$ for H11, and +13.0~km~s$^{-1}$ for H13d.

In addition to the radial velocity measurements, which are based on the 
cross-correlation of all features in the spectrum, we also assign a spectral quality 
index to each object based on what spectral features are obvious/visible relative to 
the noise.  Spectra with two or more definite absorption/emission 
lines were assigned a spectral quality index of $Q=4$, while spectra with one 
definite and one marginal line were assigned $Q=3$.  Spectra for which a radial 
velocity could not be determined due to poor S/N and/or spectra that showed a lack 
of definite features were assigned $Q=2$ (255 objects).  Spectra with extremely 
poor S/N were assigned $Q=1$, and spectra affected by instrumental artifacts 
(bad columns, CCD edges, vignetting, etc.) were assigned $Q=-2$.  Of the 
830~spectra obtained on these 6~masks, only 2~objects were assigned $Q=-2$ 
and 4~objects were assigned $Q=1$.  These objects will be ignored from here on.

\begin{figure}
\begin{center}
\leavevmode
\includegraphics[width=9cm]{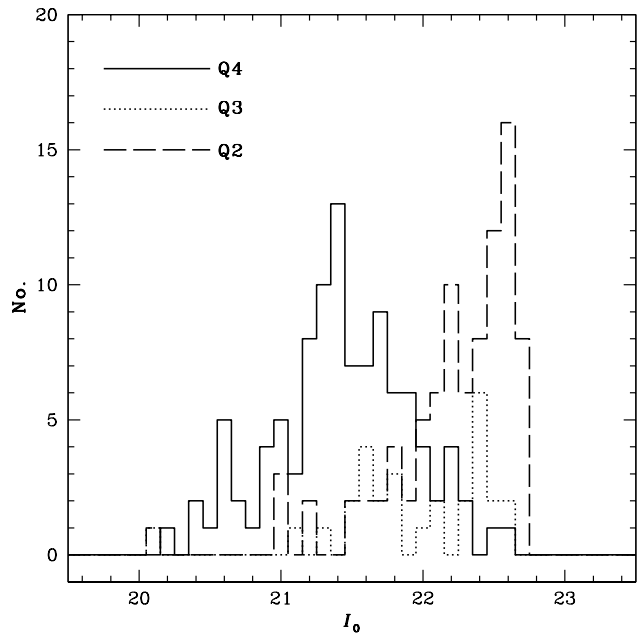}
\end{center}
\caption{Distribution of $I_0$ magnitude for spectroscopic targets in the H13s
stream field grouped according to their spectral quality codes: $Q=2$, 3,
and~4.  The distributions confirm that the faintest objects in our
spectroscopic sample ($I_0\gtrsim22$) are generally the ones for which the
radial velocity measurement fails presumably as a result of their poor spectral
quality (i.e., relatively low S/N).}
\label{fig:quality}
\end{figure}

Figure~\ref{fig:quality} presents the distribution of $I_0$ magnitudes for
objects on the two~H13s masks with spectral quality index $Q=2$, 3, and~4.  The
distribution of $Q=4$ objects is clearly shifted towards brighter apparent
magnitudes than the rest.  The $Q=2$ category comprises most of the faint
spectroscopic targets.  The bulk of the $Q=3$ objects lie between the other 
two~categories.  This figure clearly indicates that our spectral quality indices 
are correlated with the S/N in our data (i.e., a large number of stars have 
been classified as $Q=2$ precisely because they are fainter and have 
lower S/N than the rest of the sample).

\begin{figure*}
\begin{center}
\leavevmode
\includegraphics[width=12cm,angle=270]{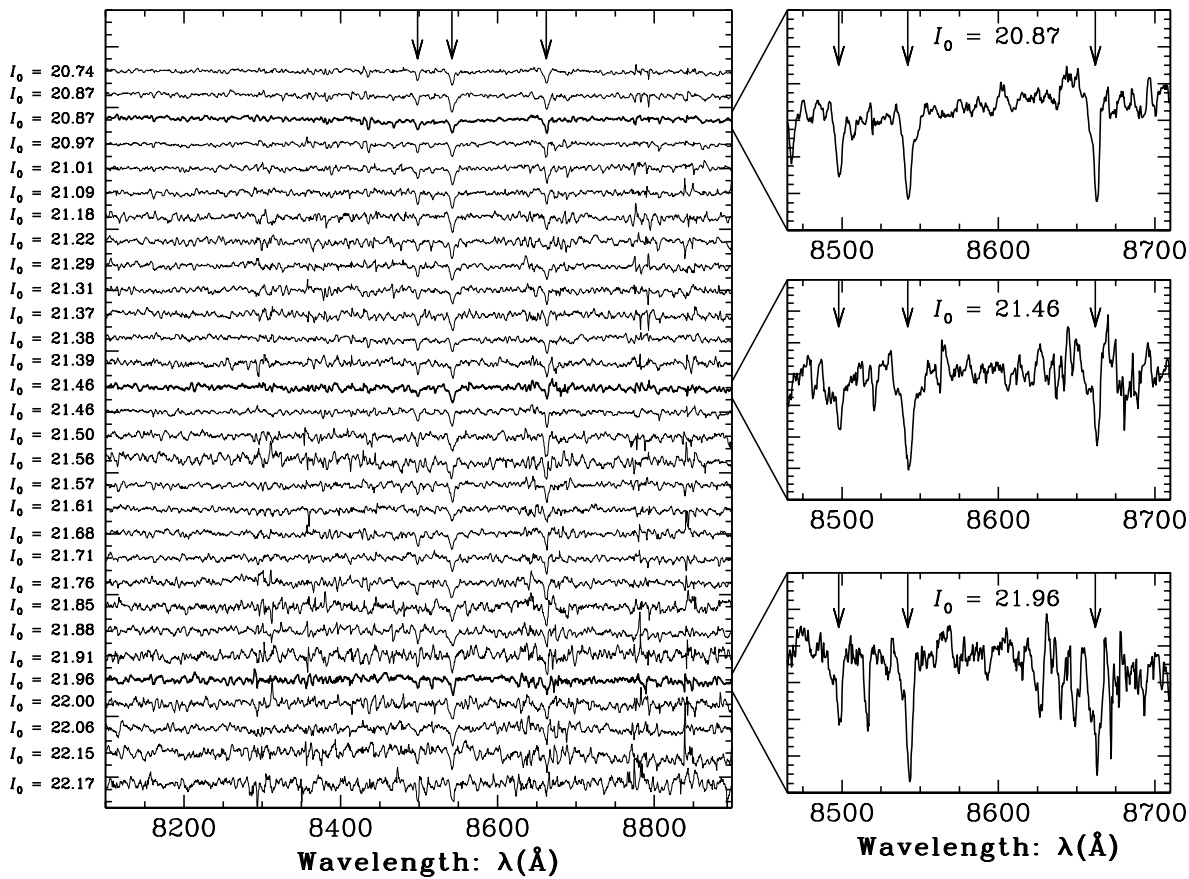}
\end{center}
\caption{Spectra for 30~M31 stream RGB stars, with $Q=4$, in the brightness range
$20.7\lesssim{I_0}\lesssim22.2$ showing only the wavelength range
8100--$\rm8900\AA$.  The spectra have been corrected to zero velocity,
normalized at $\rm\lambda\sim8500\AA$, offset in $y$ arbitrarily, and smoothed
using a 10~pixel ($\rm3\AA$) boxcar function for illustration purposes only.
Arrows mark the locations of the Ca\,{\sc ii} triplet: 8498, 8542, and
$\rm8662\AA$.  The panels on the right present a closer look at the Ca\,{\sc
ii} triplet region for three arbitrarily chosen stars.}
\label{fig:spectra}
\end{figure*}

In Figure~\ref{fig:spectra} (left) we present spectra of 30~M31 RGB stars in
the H13s stream field, displaying an $\rm800\AA$ portion in the region of the
Ca\,{\sc ii} triplet: 8498, 8542, and $\rm8662\AA$ (the full spectra cover a
$\rm2700\AA$ range, centered on $\rm\approx7800\AA$).  The stars whose spectra
are shown in this plot span the entire brightness range over which we have
assigned a spectral quality index of $Q=4$ (see Fig.~\ref{fig:quality}).  In 
Figure~\ref{fig:spectra} (right) we zoom into the Ca\,{\sc ii} triplet region for 
three arbitrarily chosen stars with $I_0=20.87$, 21.46, and 21.96.  At this scale, 
it is clear that the spectra also show several other absorption features within this 
narrow wavelength window, such as Fe\,{\sc i} lines at $\rm\sim8517\AA$ and 
$\rm\sim8690\AA$ (see middle and bottom right panels).

\section{Removing Contaminants} \label{cleansample}

The line of sight to M31 encounters both foreground and background contamination.
Although extragalactic objects can be removed from image morphology and radial velocity 
alone, a clear study of M31's stellar populations also requires the removal of Milky 
Way disk/halo stars.  In radial velocity space, these dwarfs often overlap M31 giants and 
can therefore be indistinguishable.  For example, our H13d pointing encounters M31's disk 
in the north-east quadrant, where the disk is redshifted relative to M31's $\sim$$\rm-$300~km~s$^{-1}$ 
blueshift.  Therefore, these stars will have low radial velocities, similar to those for 
Galactic dwarfs.

We have developed a sensitive technique to eliminate Milky Way contaminants from our study of M31 stars.  
The method uses probability distribution functions calculated from a training set of known 
RGB and dwarf stars.  Using four criteria: (1)~radial velocity, (2)~Na\,{\sc i} equivalent width 
(surface gravity sensitive), (3)~position in the CMD, and (4)~comparison between photometric 
and spectroscopic [Fe/H] measurements, each star is assigned four likelihood values of being a 
giant or a dwarf based on its value within each diagnostic.  The details of this procedure 
are given in Gilbert et al.\ (2005, in preparation), along with one other diagnostic that 
was not available for these fields ($DDO51$ parameter).  Here we provide a few details and present 
the separation of RGB from dwarf stars in just the stream field, H13s.  

\begin{figure*}
\begin{center}
\leavevmode
\includegraphics[width=14cm]{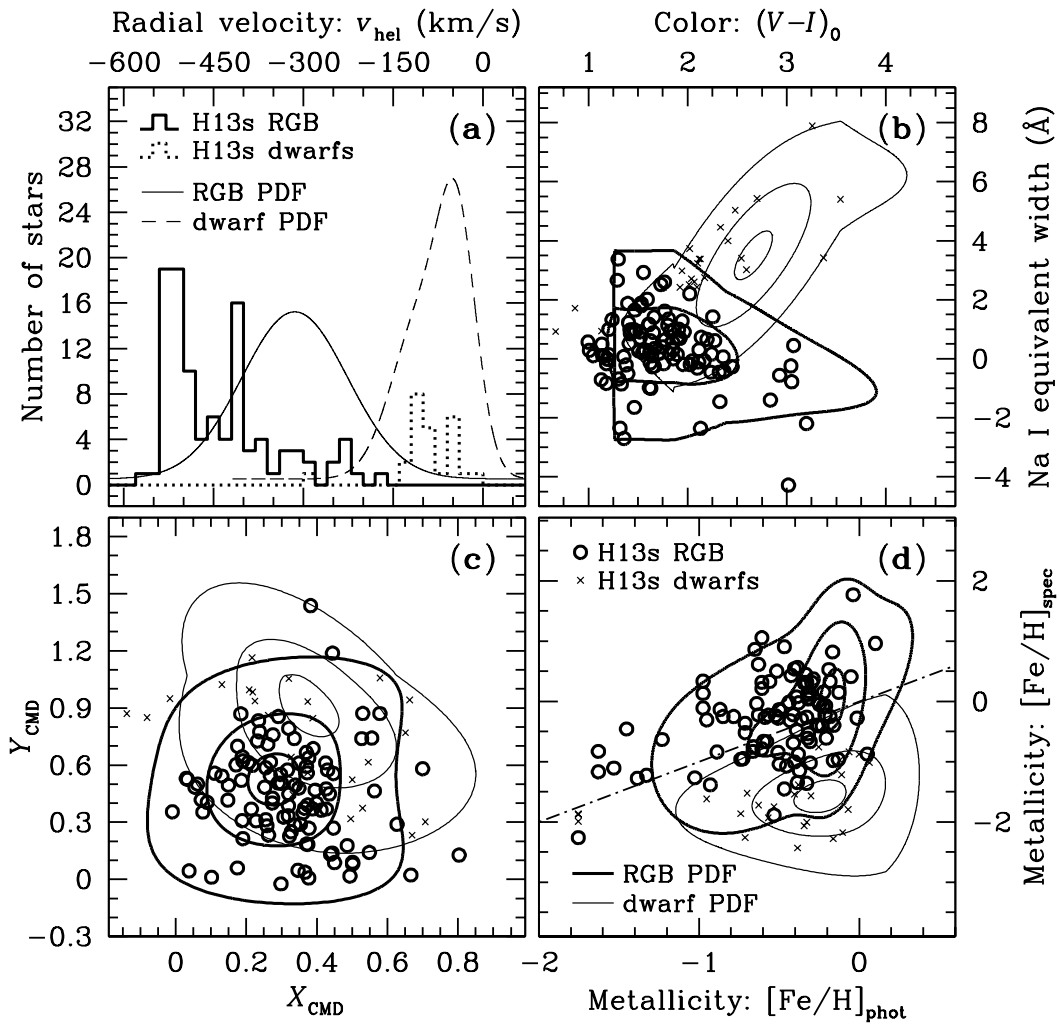}
\end{center}
\caption{Four of the photometric/spectral diagnostics that are used to
distinguish M31 RGB stars from foreground Milky Way dwarf contaminants 
are shown.  The probability distribution functions for the training sets 
are given as darker solid curves for RGB stars and dashed (or thinner curves) 
for the dwarfs, in each of the 4 panels: (a) radial velocity, (b) Na\,{\sc i} 
equivalent width, (c) position in the CMD, and (d) comparison between 
photometric and spectroscopic [Fe/H] measurements.  Also shown are 
the positions of H13s stars within each diagnostic.  The computed 
RGB stars are shown as a solid histogram in (a) and as large open circles 
in (b), (c), and (d).  Similarly, the computed dwarf stars in H13s are shown 
as a dashed histogram in (a) and as small crosses in (b), (c), and (d).  In 
all panels, the confirmed RGB and dwarf stars appear to follow the appropriate 
probability distribution functions.  Note that the RGB velocity histogram in 
(a) is offset from the training set distribution due to the large negative 
velocity of stream stars relative to M31's systemic velocity.}
\label{fig:h13s4pan}
\end{figure*}

In Figure~\ref{fig:h13s4pan}, the probability distribution functions for each 
of the four diagnostics discussed above are given.  The darker solid curves denote the RGB 
training set population 
whereas the dashed or thinner curves represent the dwarf training set.  Clearly, the probability 
distribution functions have different shapes although there is some overlap.  The probability 
of any given star in our study being a RGB and a dwarf is calculated given its properties 
within each of these diagnostics and the corresponding probability distribution 
functions.  In this Figure, we also overplot the H13s stars within each 
diagnostic.  For $v_{\rm hel}$, the H13s stars are shown as histograms for calculated 
RGB stars (solid) and dwarfs (dashed).  The H13s RGB histogram is offset from the training 
set RGB curve given the large negative velocities of stream stars in this field (our training 
set is drawn from mostly RGB stars in M31's spheroid).  For the other three diagnostics, the RGB stars 
are shown as larger open circles and the dwarfs are shown as smaller crosses.  In all panels, the 
confirmed RGB and dwarf stars follow the appropriate probability distribution function.

\begin{figure}
\begin{center}
\leavevmode
\includegraphics[width=9cm]{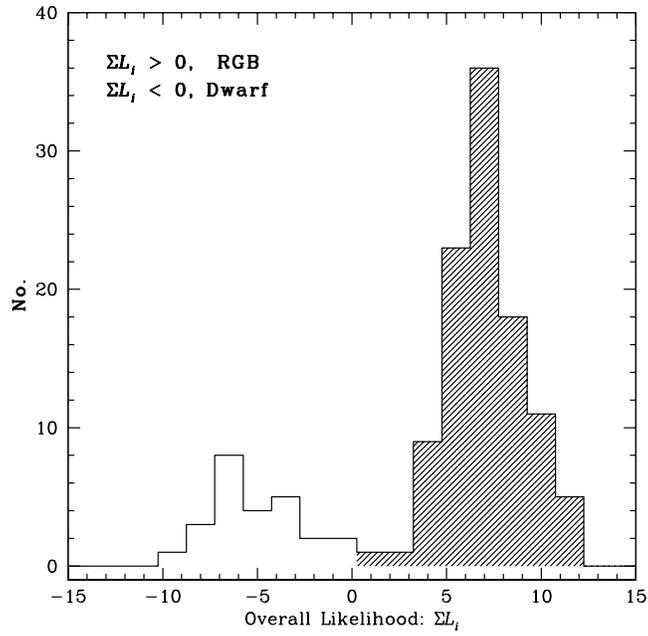}
\end{center}
\caption{The individual probabilities of a star being an RGB or a dwarf from the four 
diagnostics in Figure~\ref{fig:h13s4pan} are combined to yield an overall likelihood, 
$\Sigma{L_i}$.  The results show a very nice bi-modal distribution with a minimum 
near 0.  The shaded histogram ($\Sigma{L_i} >$ 0) represents the RGB stars, whereas 
the open histogram ($\Sigma{L_i} <$ 0) shows the dwarfs.}
\label{fig:h13slikelihood}
\end{figure}

As discussed in Gilbert et al.\ (2005, in preparation), the individual probabilities from 
each of these four diagnostics are combined to yield the final discriminant of whether 
the star is a bona-fide M31 giant.  This overall likelihood ($\Sigma{L_i}$), for H13s stars, 
is presented in Figure~\ref{fig:h13slikelihood}.  The results show a very nice bi-modal 
distribution with RGB stars ($\Sigma{L_i} >$ 0) and dwarf stars ($\Sigma{L_i} <$ 0) well 
separated.  Therefore, although none of the individual diagnostics used to measure 
the probability distribution functions provide a perfect discriminant of RGB stars and dwarfs 
(with the exception of radial velocity, in the rare case of stream stars), the combination 
of the four diagnostics provides a powerful tool.

The remainder of our analysis focuses on a sample of M31 RGB stars cleaned by the 
above process.

\section{Imaging/Photometric Analysis} \label{imageanal}

\subsection{Color-Magnitude Diagrams} \label{CMDs}

\begin{figure*}
\begin{center}
\leavevmode
\includegraphics[width=14cm]{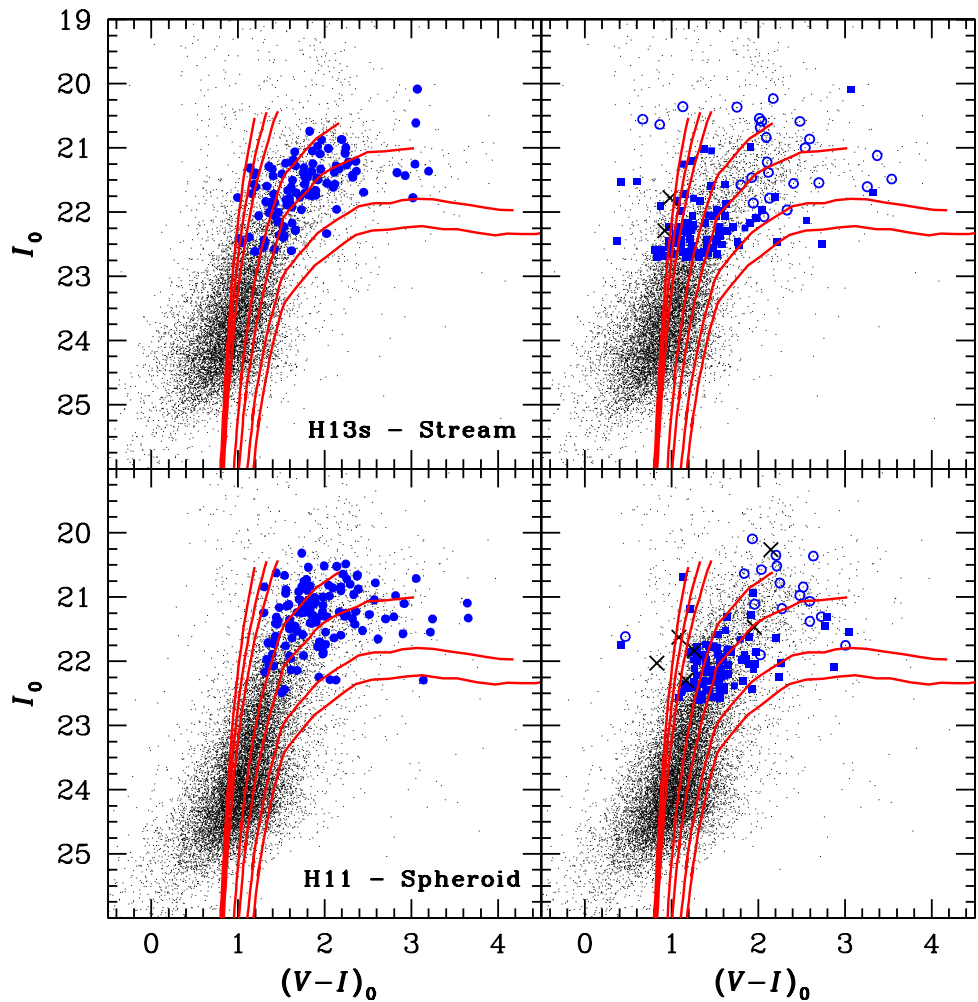}
\end{center}
\caption{Color-magnitude diagrams for the H13s (top) and H11 (bottom) fields.
Bonafide M31 RGB stars (i.e., those that have spectral quality $Q=3$ or~4 and
have passed the foreground dwarf star contaminant screening in
\S\,\ref{cleansample}) are shown in the left-hand panels as large filled
circles.  The right-hand panels show Milky Way dwarf stars (open circles),
$Q=2$ objects for which there is no reliable radial velocity measurement
(filled squares), and background field galaxies (crosses).  The smaller dots
represent stars in a larger imaging region surrounding the spectroscopic
fields.  A detailed discussion of these data and isochrones is given in
\S\,\ref{CMDs}.}
\label{fig:cmdall}
\end{figure*}

In Figure~\ref{fig:cmdall}, we present CMDs for each of the H13s (top) and H11 (bottom) fields.  
The objects for which we obtained a spectra are shown as larger symbols in either panel.  
These were selected to be bright enough for good S/N in one hour exposures with Keck, and therefore 
have $20.5\lesssim{I_0}\lesssim22.5$.  The remaining objects (shown as smaller dots) represent 
the CFHT photometry for a very small section of the MegaCam image, equal to approximately 
1 out of the 36 chips on the mosaic ($\approx96$~arcmin$^2$), centered at the DEIMOS pointing.  
In the CFHT photometry for H13s, there are a total of $\sim2400$ stars with
$20\lesssim{I_0}\lesssim22.5$.  Scaling this to the size of the DEIMOS
footprint gives $\sim2600$~objects.  This 
is calculated using the area of two masks ($16'\times4'$ each), and
subtracting the area of the overlap region (23.9~arcmin$^2$).   Of these, 270
(10\%) were targeted for spectroscopy in our two masks.

Reddening corrections have been applied to the data using the Schlegel, Finkbeiner, \& Davis 
(1998) reddening maps.  For H13s, 
the correction is found to be $E(V-I)=0.08$ and for H11 it is $E(V-I)=0.10$.  Each of 
the CMDs clearly shows the RGB of M31 extending from the faintest stars ($I_0\sim25$) up to 
the tip of the RGB at $I_0\sim20.5$.  For clarity, we have also overplotted several theoretical 
isochrones from Girardi et~al.\ (2002) ranging in metallicity from $Z=0.0001$
or $\rm[Fe/H]=-2.3$ (bluest isochrone) to $Z=0.03$ or $\rm[Fe/H]=+0.2$ (reddest isochrone) for
$t=12.6$~Gyr.  A distance modulus of $(m-M)_0=24.47$, corresponding to a distance of 783~kpc, 
has been applied to the isochrones.

The left panels of Figure~\ref{fig:cmdall} show the cleaned sample of
confirmed M31 RGB stars in 
the stream and spheroid (i.e., those that have passed a $Q=3$ or 4 cut, and survived the 
dwarf star contamination rejection routines discussed in \S\,\ref{cleansample}).  The panels 
on the right show confirmed Milky Way dwarfs (open circles), $Q=2$ objects (filled squares), 
and galaxies (crosses) within each of these fields.  As expected, most of the Milky Way 
dwarfs are located above the tip of the RGB and most of our $Q=2$ objects represent the 
faintest stars for which we obtained a spectrum.  The cleaned sample of stream stars (top-left) 
also shows a few confirmed M31 giants that are above the tip of the RGB.  These are potentially 
intermediate aged asymptotic giant branch (AGB) stars.  Only minor differences exist between the 
metallicity distributions of stars in these fields.  As discussed in \S\,\ref{masks}, 
the selection process involved picking stars within a magnitude 
range and did not bias the color selection.  The majority of the stars in the spheroid field 
appear to have a similar metallicity and metallicity dispersion, than the stream.

\subsection{Photometric Metallicities} \label{photmet}

\begin{figure}
\begin{center}
\leavevmode
\includegraphics[width=9cm]{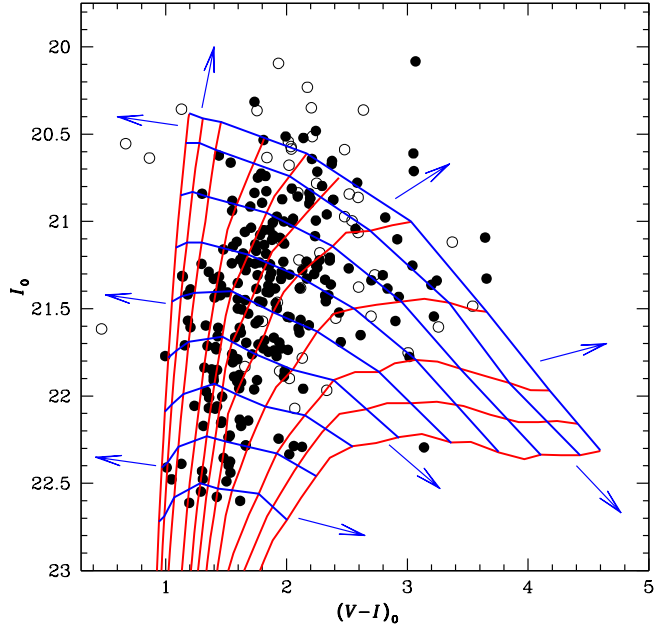}
\end{center}
\caption{A sample grid that was used to measure photometric metallicities is
shown.  Eleven representative isochrones with ages of 12.6~Gyr and
$0.0001<Z<0.03$ or $\rm-2.3<[Fe/H]<+0.2$ (Girardi et~al.\ 2002) are overlaid on
all confirmed RGB stars in the H13s and H11 fields (filled circles).  Milky Way
dwarf stars are shown as open circles.  ``Cross-isochrones'' are also shown as
discussed in \S\,\ref{photmet}.  Photometric metallicities of all stars were
measured by using a nearest neighbor approach within a much more finely spaced
grid of 300~isochrones.  Arrows indicate the extrapolation directions.}
\label{fig:cross}
\end{figure}

In order to compute photometric metallicities for these stars, we used a very 
large grid of finely spaced ($\Delta{Z}=0.0001$) isochrones of age 12.6 Gyrs 
and $\rm[\alpha/Fe]=0$ (L.\ Girardi 2004, private communication). The metallicity of 
each star was derived using a nearest neighbor method within this grid.  The most metal-poor 
isochrone used was $Z=0.0001$ ($\rm[Fe/H]=-2.3$) while the most metal-rich 
isochrone had $Z=0.03$ ($\rm[Fe/H]=+0.2$), resulting in over 300~isochrones and 
15,000~data points within the region of interest in the CMD.  We used linear 
extrapolation to ``pad'' the isochrones to accommodate stars that lie beyond the 
bounds of the models in the CMD.  For the brighter data points, the standard 
isochrones were extrapolated brightward and for the bluer/redder data points, we produced 
``cross-isochrones'' by connecting the models with different $Z$, yet the same mass, 
and extrapolated those (see Fig.~\ref{fig:cross}).  These ``cross-isochrones'' were also 
used in the separation of Milky Way dwarfs from M31 giants as discussed in \S\,3.5 of 
Gilbert et~al.\ (2005, in preparation).  As can be seen from Figure~\ref{fig:cross}, this 
extrapolation was needed for very few stars thus minimizing any possible contamination 
from treating an AGB star as a RGB star.

\begin{figure}
\begin{center}
\leavevmode
\includegraphics[width=9cm]{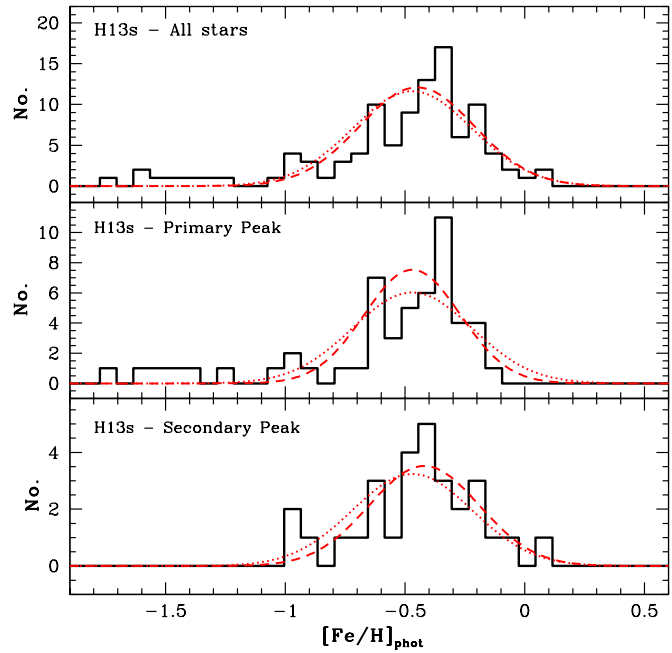}
\end{center}
\caption{The photometric metallicity distributions for H13s is shown.  The top panel 
presents the data for all M31 RGB stars in the field, whereas the middle and bottom 
panels present only the data in the primary and secondary velocity peaks (to be discussed in 
\S\,\ref{kinematics}).  The dashed curves show Gaussian fits to each population.  For 
the entire sample (top), we find $\rm\langle[Fe/H]\rangle=-0.45$, $\rm\sigma([Fe/H])=0.24$.  
The dotted curves shows the metallicity distribution of stars in our spheroid field, H11, 
which is clearly found to be similar to our stream field.}  
\label{fig:photmet}
\end{figure}

In Figure \ref{fig:photmet} (top), we present the photometric metallicity distributions 
for stars in H13s (solid line).  For the best fit Gaussian (dashed curve), we find 
$\rm\langle[Fe/H]\rangle=-0.45$, $\rm\sigma([Fe/H])=0.24$.  Also shown is a Gaussian 
representing the metallicity distribution of stars in our spheroid field, H11 (dotted 
curve - Rich et al.\ 2005, in preparation).  The mean metallicity of the 
stream field is clearly similar to that of the spheroid, and with a similar dispersion.  
Guhathakurta et~al.\ (2005a) find the stream's metallicity 
to be similar ($\rm[Fe/H]=-0.54$) at a location 31~kpc from the center of M31 (i.e., 10~kpc further out).  
The metallicity dispersion in their work ($\rm\sigma([Fe/H])=0.25$) is also very similar to that 
found in this study, $\rm\sigma([Fe/H])=0.24$.  As a cautionary note, we point out that 
the relative measurement of photometric metallicity within a given population is more 
accurate than the absolute comparison of different populations.  The latter will be 
more affected by several uncertainties: possible age spreads between the 
populations, [$\alpha$/Fe] variations, differential reddening, etc.  The middle and 
bottom panels of Figure \ref{fig:photmet} will be discussed in \S\,\ref{interpstream}.

\section{Spectroscopic Analysis} \label{spectraanal}

\subsection{Spectroscopic Metallicities} \label{specmet}

We derive spectroscopic metallicities for RGB stars in this work based on the line 
strength of the Ca\,{\sc ii} triplet ($\rm\lambda\sim8500\AA$).  These are computed on 
the Carretta \& Gratton (1997) scale.  This procedure is described in detail in Reitzel \& 
Guhathakurta (2002) and Guhathakurta et~al.\ (2005a).  Summarizing, the equivalent 
widths of the three individual Ca\,{\sc ii} lines are measured and combined to 
yield a reduced equivalent width according to Rutledge, Hesser, \& Stetson  (1997).  
An empirical calibration based on Galactic globular cluster RGB stars was used to 
calculate [Fe/H]$_{\rm spec}$ from the equivalent widths (Rutledge et al.\ 1997).  
A value of $V_{\rm HB}$ = 25.17 (Holland 1996) was adopted for the 
luminosity-based correction of surface gravity.  This comes from {\it HST\/} observations 
of M31 HB in nearby fields.  This measurement, from equivalent widths, is known to 
have a relatively large uncertainty in it compared to the random error in [Fe/H]$_{\rm phot}$, 
yielding an inflated spread in metallicities.  The two however are consistent on a 
global scale, as shown in \S\,5.2 of Guhathakurta et~al.\ (2005a).

\subsection{Kinematics} \label{kinematics}

\begin{figure}
\begin{center}
\leavevmode
\includegraphics[width=9cm]{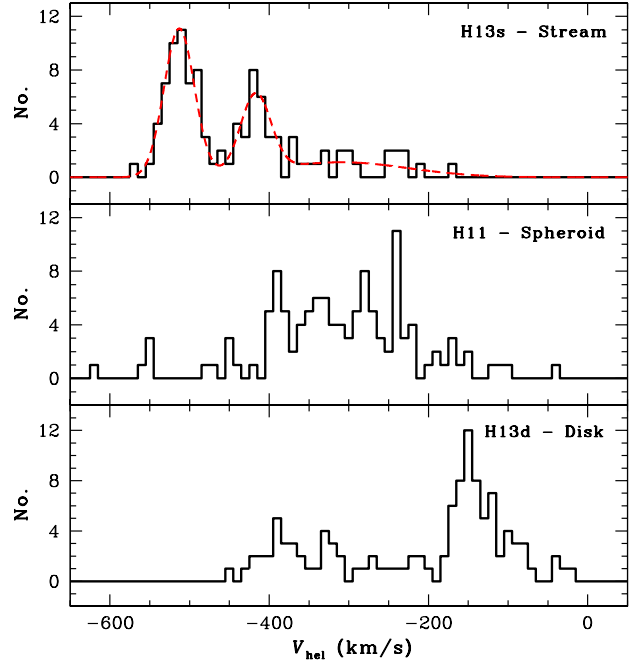}
\end{center}
\caption{The velocity histograms for our three fields show a very clean separation of the 
stream, spheroid and disk of M31.  The stream component contains 
a dynamically cold population of highly blueshifted stars, $\bar{v_1}=-513$~km~s$^{-1}$, the spheroid 
contains a hotter population of stars moving at M31's systemic velocity $v_{\rm 
hel}\sim-300$~km~s$^{-1}$, and the disk in the north-east quadrant shows a cold population of 
low velocity stars (as expected given the disk's rotation in this quadrant).  Also shown is a 
triple Gaussian fit to the two prominent stream peaks, as well as the underlying smooth spheroid, 
in the top panel (see \S\,\ref{kinematics}).}  
\label{fig:newcombvel}
\end{figure}

As discussed in \S \ref{velocityquality}, we measure velocities for all stars by 
cross-correlating the observed spectrum with template spectra over a range of 
velocities.  In Figure~\ref{fig:newcombvel} we present velocity histograms for M31 
stars in H13s, H11, and H13d.  The histograms show a clear separation of M31's stellar 
populations, based solely on kinematics.  As expected, the stream field at 
$R\sim21$~kpc shows a large number of highly blueshifted stars as well as some 
evidence of a weakly detected smooth spheroid.  We also see a distinct, cold secondary 
peak of M31 stars at a slightly less negative velocity.  Such a bimodal velocity distribution 
has never been reported before and may contain important information related to the orbit of the 
giant southern stream, or the discovery of a new stream (see discussion below).  The H11 minor-axis 
field shows a much hotter population of stars (presumably spheroid) centered near M31's 
systemic velocity.  The H13d major-axis field in 
the north-east quadrant shows a dominant cold population superimposed on a smooth spheroid.  
The velocity of this cold component is generally consistent with M31's disk rotation in 
this field (i.e., the disk is rotating away from the line of sight in this quadrant 
causing a relative redshift of stars on top of the $\sim$$\rm-$300~km~s$^{-1}$ systemic 
blueshift).  

We focus now on the kinematics of the H13s population shown in Figure~\ref{fig:newcombvel} (top).  
The velocity histogram is well fit by the sum of three Gaussians (dotted line): the two prominent 
cold peaks in velocity space, and the broad underlying smooth component.  For the 
best fit, we find the mean velocity of stars in the primary peak is $\bar{v_1}=-513$~km~s$^{-1}$ 
and $\sigma(v_1)=19$~km~s$^{-1}$.  For the secondary peak, we find $\bar{v_2}=-417$~km~s$^{-1}$ and 
$\sigma(v_2)=19$~km~s$^{-1}$.  The Gaussian representing the smooth spheroid of M31 is 
held fixed at $\bar{v_3}=-310$~km~s$^{-1}$ ($\sigma(v_3)=85$~km~s$^{-1}$), as constrained from 
our H11 field.  

The dispersion quoted above includes the intrinsic dispersion of the population 
and our measurement error.  We estimate the latter by comparing the velocity measurements of 
stars that were observed twice, largely in the H13d field.  The brightnesses and S/N of those 
spectra are virtually identical to the H13s stars, and therefore the measurement 
errors will be similar for both.  For the mean S/N acquired (8.8 in H13s), 
we find that the velocity measurement error is 10.89~km~s$^{-1}$.  Subtracting this 
in quadrature from the measured $\sigma(v_1)$ and $\sigma(v_2)$ above, we find that the 
velocity dispersions of stars is 16~km~s$^{-1}$ in the primary and 
secondary peaks.

Of the 104~stars in the H13s radial velocity histogram, the triple Gaussian fit to the cumulative 
distribution suggests that 50.5\% of the stars are in the primary narrow peak corresponding 
to the giant southern stream (53~stars), 26.5\% are in a second narrow peak (28~stars), and 
23\% are in the broad spheroidal component (24~stars).  The contamination of spheroid 
stars into either of our primary or secondary peaks is negligible.  Guhathakurta 
et~al.\ (2005b) compute a surface brightness from the spheroid component in H13s 
and find that these data agree well with the known M31 surface brightness profile.  
In the next section, we consider the properties of the primary and secondary 
cold peaks.  We note that the CMD positions of stars in both peaks in the CMD 
are indistinguishable. 

\section{Discussion} \label{discdisc}

\subsection{Giant Southern Stream} \label{discGSS}

As described in \S\,\ref{previous} the velocity gradient of the giant 
stream has now been mapped in several small pointings over a 125~kpc 
radial extent (Figure \ref{fig:fields} shows the approximate locations of 
these pointings).  With respect to these pointings, our H13s field is located 
very close to the center of M31 ($R \sim$ 21 kpc).  Within this radius, 
only one study has kinematically confirmed stream members (Ibata et~al.\ 
2004).  Their field~8, located at $R\sim12$~kpc), contains only four~stars 
believed to be stream members, all moving with large negative velocities, 
$-480\lesssim{v}_{\rm hel}\lesssim-560$~km~s$^{-1}$.  The distance of the stream 
in this field is found to be virtually identical to the distance of M31 
(McConnachie et~al.\ 2003).

In the stream field, our velocity histogram contains 54~stars with $v_{\rm 
hel}\leq-460$~km~s$^{-1}$ in the primary cold population, 40 of which have 
$v_{\rm hel}\leq-500$~km~s$^{-1}$.  All previous studies of M31's giant
southern stream combined have only found five~stars with $v_{\rm
hel}\leq-500$~km~s$^{-1}$ (1~star in mask \#1 of Guhathakurta et~al.\ 2005a and
1 and 3~stars in fields~6 and~8 of Ibata et~al.\ 2004, respectively).  Given
these statistics, we are investigating a largely unexplored region of radial velocity 
space for the giant southern stream. 

In fitting the giant southern stream to models, Fardal et~al.\ (2005) do not use 
the inner Ibata et~al.\ (2004) pointing noting that the data may be largely contaminated 
by M31 disk stars.  Therefore, the present study represents an important check on their 
models for the orbit of the stream at small radii from the nucleus of M31.  For a distance 
along the stream of 1.6 degrees (the location of our H13s field), their simplest model 
(radial period of 1.7 Gyrs, apocenter 119 kpc, and pericenter 2.2 kpc) predicts that 
the velocity of the stream should be $\sim-510$~km~s$^{-1}$.  Thus, their prediction 
is in excellent agreement with our measured value of $\bar{v_1}=-513$~km~s$^{-1}$.  
Other orbits, and further details for this orbit, are given in Fardal et~al.\ (2005)
  


%

\subsection{Nature of the Secondary Cold Component}
\label{interpretation}

As mentioned earlier, the H13s velocity histogram shown in Figure \ref{fig:newcombvel} shows two 
prominent kinematically cold populations of stars.  We now consider the second population at 
$\bar{v_2}=-417$~km~s$^{-1}$, both from the standpoint of it being a new stream or a part 
of/associated with the giant southern stream or, alternatively, being related to M31's outer disk. 

\subsubsection{Satellite Debris Trail?} \label{interpstream}

Studies of the Milky Way have shown that multiple streams can exist along a single line of sight 
(e.g., Newberg et al. 2002).  Numerical simulations (e.g., Bullock, Kravtsov, \& Weinberg 2001) also 
predict large numbers of halo streamers, some of which may overlap.   Disentangling this substructure 
represents an important step in quantifying the number density of streams and therefore testing 
CDM and hierarchical clustering scenarios for structure formation.  The secondary velocity peak 
in our H13s field identified above may represent a new halo tidal stream, superimposed on the giant 
southern stream.  Such a stream has not been detected in star-count maps of M31.  However, this is not surprising 
given the close location of our field to the disk of M31 which produces large overdensities in the 
star-count map that could mask any underlying substructure from a stream.  The locations of the 
29~stars in the secondary peak are distributed smoothly over the entire DEIMOS masks, suggesting 
that the spatial scale of this population is larger than the size of a mask ($\sim$16$'$).

An alternative suggestion is that the secondary peak represents a wrapped around component of the 
giant southern stream itself.  If confirmed as a stream component, the secondary velocity peak 
in our data could set very important constraints on the orbit of the stream and the nature of 
the progenitor.  As we discussed in \S\,\ref{previous}, several groups have 
constructed dynamical models to better understand the nature of the Andromeda stream and its 
progenitor.  These models make testable predictions for where debris from the progenitor 
may lie.  Although, the wide range of models presented by these groups (see Font et~al.\ 2005; 
Fardal et al.\ 2005 for details) do contain models that show a progenitor 
orbit that passes near the location of H13s twice (see Fig.~8 of Fardal et~al.\ 2005), during 
each such passage the stars are predicted to be moving in very different phases.  This results in 
large velocity offsets during each passage and therefore are inconsistent with the present 
observations.  Similarly, the best two orbital fits of Ibata et~al.\ (2004) should not produce 
the secondary peak at its observed location.  In that model, the stream reaches its maximum 
velocity on the north-west side of M31 and then falls back to a location south of M31's disk.  
The predicted velocity of the stream during its fall back is estimated to be $\sim-300$~km~s$^{-1}$ 
near the projected location of H13s on the north-south plane as defined by Ibata et~al.  However, 
such an orbit (see Figure~4(d) of Ibata et~al.\ 2004) predicts that the secondary population would 
lie almost 2 degrees east of the giant southern stream and therefore can not explain these data.  

We note that the stream's very small pericenter distance makes the post-pericentric portion of the orbit 
strongly dependent on the (uncertain) details of M31's inner potential (Geehan et~al.\ 2005).  
The test particle orbit approximation used in the simulations above is clearly an oversimplification.  
More realistic modelling using $N$-body satellites (e.g., Fardal et~al.\ 2005) 
shows that although the secondary peak in these data can not be reproduced by the stream's orbit, 
reasonable orbits can be derived that connect several other prominent features/satellites of 
M31 with the stream: Northern Spur, M32, And~NE, and a concentration in the planetary nebulae 
system discovered by Merrett et~al.\ (2003).  Connecting these features with the 
stream is possible given that the phase of the progenitor is unknown 
(i.e., M32 has an inconsistent radial velocity with stream stars in its present 
location, however, its present location could be one radial orbit ahead of the visible 
debris).  Better orbital calculations to narrow down the progenitor and orbit of the stream 
require more observational constraints, especially in the locations close to M31 
where the stream's orbit is not well constrained. 

\begin{figure}
\begin{center}
\leavevmode
\includegraphics[width=9cm]{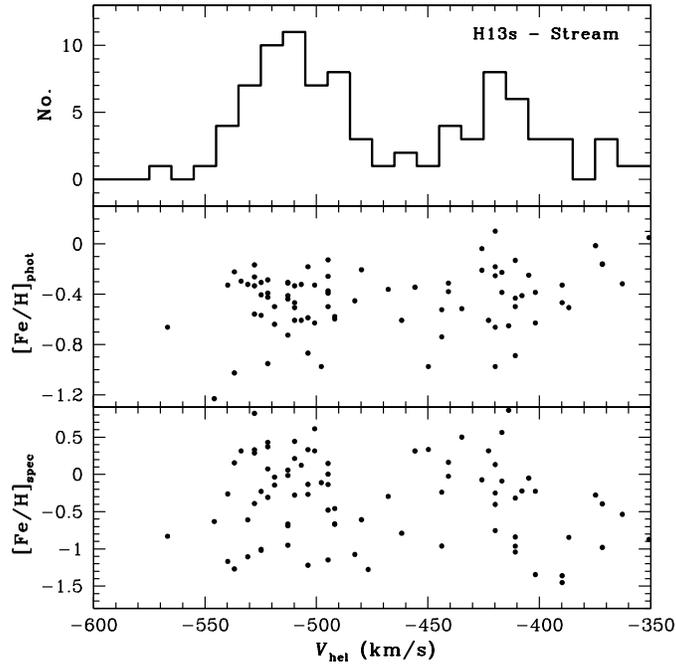}
\end{center}
\caption{The velocity histogram, photometric metallicity vs. velocity, and spectroscopic 
metallicity vs. velocity distributions for the bimodal velocity peaks in the stream field. 
The metallicities of the two populations in the dynamically cold peaks are clearly similar 
(see \S\,\ref{interpretation} for a discussion).  The larger spread in the spectroscopic 
metallicity distribution is caused by the propagation of the relatively large
random errors in the Ca\,{\sc ii} triplet equivalent 
width measurements.  Only stars with $v_{\rm hel}\leq-350$~km~s$^{-1}$ are shown.}
\label{fig:streammet}
\end{figure}

We can also address the nature of the secondary peak in our velocity histogram, and whether 
or not it may be related to the stream, by investigating the chemical composition 
of stars in the peak.  In Figure~\ref{fig:streammet}, we present a three-panel view of 
the velocity histogram (top), photometric [Fe/H] estimate vs.\ radial velocity (middle), and
spectroscopic [Fe/H] estimate vs.\ radial velocity (bottom) for the stream field
H13s.  The middle and bottom panel of this diagram indicate that the mean metallicity, 
and metallicity dispersion, of the populations in the primary and secondary peaks are 
very similar.  The only difference in the photometric metallicity distribution of stars in the 
two peaks is a small tail with higher metallicities in the secondary population that is lacking 
in the primary.  This difference is subtle, constrained by only a few stars in the 
photometric metallicity diagram and is not seen in the spectroscopic metallicity distribution.

In Figure \ref{fig:photmet}, we fit the metallicity distributions of 
these two kinematically distinct populations to Gaussians and find that indeed the 
distributions are very similar.  Specifically, for the primary component in our 
velocity histogram we find $\rm\langle[Fe/H]\rangle=-0.47$ ($\rm\sigma([Fe/H])=0.20$) and 
for the secondary peak we find $\rm\langle[Fe/H]\rangle=-0.42$ ($\rm\sigma([Fe/H])=0.23$).  
Although the similar nature of the metallicity distributions in these two peaks naively 
suggests a similar origin for the stars, we note that the mean metallicity and
metallicity dispersion of stars in the disk of M31 are also similar to those of
the stream (Reitzel et~al.\ 2005, in preparation).  Therefore, from these data alone, 
it is equally likely that the secondary peak represents some component of the disk of 
M31 (e.g., a warp or an overdensity induced by satellite interaction --- see 
\S\,\ref{interpdisk}).

Another possible variation could be that the observed secondary velocity peak does not necessarily 
represent a different part of the orbit of the giant stream, but rather some unbound debris that is 
orbiting differentially from the main body of the stream.  Future simulations with added constraints 
from these data and future observations (see \S\,\ref{future}) will explore these possibilities and 
likely shed light on any relation of this population to the giant southern stream.

\subsubsection{M31 Disk Feature?} \label{interpdisk}

We noted earlier that the metallicity distribution of stars in the secondary 
peak are similar to the disk field H13d (although they are at odds with the metallicity 
reported for the extended disk of M31 by Ibata et~al.\ 2005).  The Ibata 
et~al.\ (2005) star-count map (see their Fig.~1) shows several diffuse 
features surrounding M31's disk 
and so it is possible that a kinematically cold population of metal-rich disk
stars (or disk feature) comprise the secondary peak.  In fact, the expected
blueshift of stars in M31's disk at the location of H13s agrees well with the
location of the secondary peak in the velocity histogram.  The fits of 
Sawa \& Sofue (1981) to the H\,{\sc i} contours of Cram, Roberts, \& Whitehurst 
(1980) also suggest that the disk velocity at this location is similar to that 
measured in the secondary peak.  If the secondary peak does 
represent disk stars this would be an important constraint on determining the
extent to which the stellar disk of M31 dominates over the spheroid/halo (Worthey
et~al.\ 2005).  

To test the hypothesis that the secondary cold component represents the
smooth disk of M31, we have computed the $V$-band surface brightness that
corresponds to the stars in the secondary peak of our data (29~stars with
$-460\leq{v}_{\rm hel}\leq-380$~km~s$^{-1}$).  The surface brightness
computation is based on first using the results of \S\,\ref{cleansample} to 
isolate both confirmed M31 RGB stars and confirmed Milky Way dwarf stars.  
We note that the sampling rate and the degree of incompleteness between these 
two populations is similar in our field.  The observed ratio of these stars, 
M31 RGB to foreground Galactic dwarfs, is then multiplied by the surface 
density of Milky Way stars predicted by a Galactic star-count model 
(Ratnatunga \& Bahcall 1985).  This surface density only enters our 
final surface brightness estimate in a relative sense as we will be 
comparing the stream field to other fields, such as the H13d disk 
measurement.  Therefore, any absolute mismatch between the Galactic star 
count model and the data is not important - only variations in the degree of 
mismatch across the fields being considered.  Finally, we normalized the 
measurements to the $V$-band surface brightness estimates obtained by 
Ostheimer (2002) using a $DDO51$-selected photometric sample of M31 RGB 
candidates.  Further details of this surface brightness calculation procedure
are described in Guhathakurta et~al.\ (2005b).  

We correct the projected radius along the giant southern stream 
to a radius in the plane of the disk, $R_{\rm disk}$.  For the 
latter, we have assumed a disk isophotal flattening of $(b/a)=1-\epsilon=0.22$ 
corresponding to an inclination of $77^\circ$ (Walterbos \& Kennicutt 1988) 
and calculated $R_{\rm disk}\equiv\sqrt{a^2+[b/(1-\epsilon)]^2}$, where $a$ 
and $b$ are the projected distances (in kpc) of the field from the center 
of M31 measured along the major and minor axes, respectively.

\begin{figure}
\begin{center} \leavevmode
\includegraphics[width=9cm]{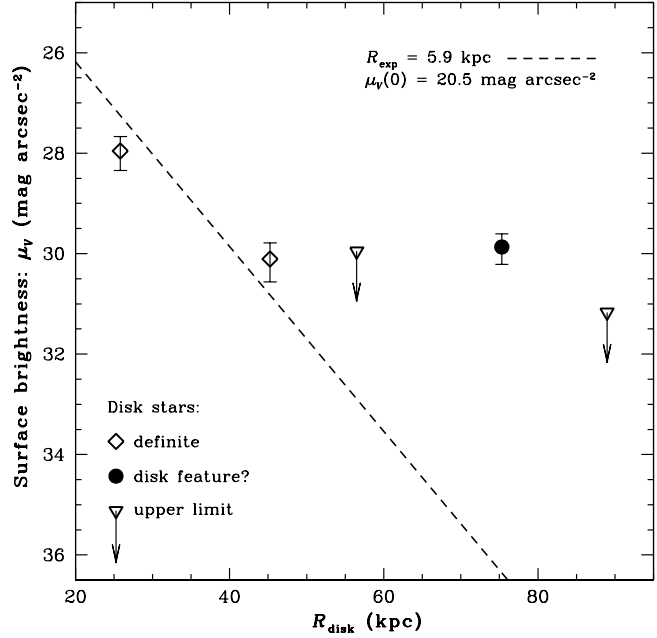}
\end{center}
\caption{Plot of the radial surface brightness profile in the $V$ band
showing the secondary component in the H13s field as a filled circle, disk
stars in the major-axis fields H13d and G1 (Reitzel et~al.\ 2004) as open 
diamonds, and upper limits to the disk population in the minor-axis fields 
H11 and RG02 (Reitzel \& Guhathakurta 2002) as open triangles with arrows.  
The error bars represent Poisson errors.  The surface brightnesses have been
calculated as discussed in \S\,~\ref{interpdisk} and Guhathakurta et~al.\
(2005b).  Also plotted is an exponential profile: $\mu_S\propto{\rm
exp}(-R_{\rm disk}/R_{\rm exp})$, where $R_{\rm exp}$ is taken to be 5.9 kpc 
(Walterbos \& Kennicutt 1988).  The H13s secondary component (filled 
circle) is clearly too bright to be related to M31's smooth disk.}
\label{fig:surfbright}
\end{figure}

In Figure~\ref{fig:surfbright}, we plot this surface brightness vs.\ $R_{\rm
disk}$ for five~fields: (1)~our H13s secondary population, (2)~disk stars in
the north-east major axis H13d field (see Fig.~\ref{fig:newcombvel} and
Reitzel et~al.\ 2005, in preparation) and south-west major axis field near G1
(Reitzel, Guhathakurta, \& Rich 2004), and (3)~rough, but conservative, upper
limits to the disk population in the south-east minor-axis fields H11 (see
Fig.~\ref{fig:newcombvel}) and RG02 
(Reitzel \& Guhathakurta 2002).  We also show an exponential disk surface
brightness profiles of the form $\mu_S\propto{\rm exp}(-R_{\rm disk}/R_{\rm
exp})$ with $R_{\rm exp}=5.9$~kpc.  This scale length is the distance corrected 
(783~kpc) $R$-band value in Walterbos \& Kennicutt's (1988) study of M31's disk.  
These results clearly show that the surface brightness of the secondary 
component of H13s (filled circle) is at least $25\times$ higher than expected 
from the extrapolation of M31's smooth exponential disk.  
This effectively rules out the smooth disk of M31 as the source of the
secondary cold population in the H13s field.  We note that Ibata et al.~(2005) find 
that the extended stellar disk is a continuation of the inner disk out to 40 kpc.  
The radius in the disk of M31 for our H13s field is almost twice as large as 
this, $\sim$75~kpc.

We can further test whether the H13s secondary component represents M31 disk
stars by considering previous studies of the stream and searching for
evidence of a secondary component in the velocity histograms.  Although these
studies have not been able to find large numbers of stars belonging to the
inner parts of the stream ($R\lesssim25$~kpc), there have been a few studies in regions
just outside this radius.  The secondary peak in these data may have been
undetected given low number statistics, or the fact that the radial velocity
of the stream becomes more positive as the distance from M31 increases and
therefore the kinematics of the stream stars may overlap with the disk stars.
In contrast, if the secondary peak in our data is M31's outer disk, then we would 
expect its velocity to be roughly constant at all radii along the stream 
($\bar{v_2}=-417$~km~s$^{-1}$).  We have shown approximate locations of three other 
studies of the giant southern stream in Figure \ref{fig:fields}, the a3 field 
(Guhathakurta et~al.\ 2005a), and the F6 and F8 fields (Ibata et~al.\ 2004).

\begin{figure}
\begin{center}
\leavevmode
\includegraphics[width=9cm]{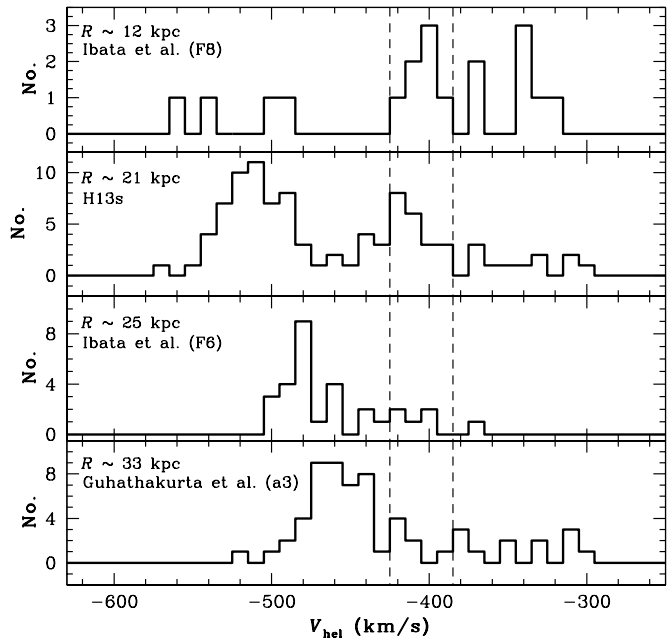}
\end{center}
\caption{Radial velocity histograms for various studies of M31's giant
southern stream are shown in order of increasing radius.  The Ibata et~al.\
(2004) fields~8 and~6 data are shown in the top and third panel, 
respectively.  Our H13s data are shown in the second panel, and the
Guhathakurta et~al.\ (2005a) $R\sim33$~kpc field~a3 is shown in the bottom
panel.  As the radial distance from M31 increases, the ratio of the number of
stars in the secondary-to-primary peak decreases (suggesting that the
secondary peak may be related to M31's outer disk --- however, see
\S\,\ref{interpdisk}).  The dashed lines indicate the region of velocity
space within which we count secondary peak stars.  Only stars with $v_{\rm
hel}\leq-300$~km~s$^{-1}$ are shown in this figure.}
\label{fig:ibata}
\end{figure}

At $R\sim33$~kpc, Guhathakurta et~al.\ (2005a) found 45 M31 stream RGB stars
centered at a velocity of $-458$~km~s$^{-1}$ (field a3).  As expected, given the further
location of their field from M31, stream stars do not have a large contrast
with the potential secondary peak.  We present their velocity histogram in
the bottom panel of Figure~\ref{fig:ibata}.  Interestingly, there is a very
small peak at $v_{\rm hel}\sim-420$~km~s$^{-1}$ in their data however it is
also consistent with the tail of the primary stream population.  Better
constraints can be derived from the $R\sim25$~kpc (field~6) Ibata et~al.\
(2004) study\footnote{Ibata et~al.\ (2004) do not provide velocity histograms
for their data so we have constructed them from their Figure~1.}.  Although
they attribute 30~stars to the stream (see their Fig.~1, right panel), only
22 of these have $v_{\rm hel}\lesssim-450$~km~s$^{-1}$.  The remaining
8~stars have $-440\lesssim{v}_{\rm hel}\lesssim-390$~km~s$^{-1}$, consistent
with the mean radial velocity of our H13s secondary peak.  These data are
presented in the third panel of Figure~\ref{fig:ibata}.  Besides our study
(second panel of Fig.~\ref{fig:ibata}), the only other spectroscopic
observations of the inner stream are the innermost field (field~8) of Ibata
et~al.\ (2004) at $R\sim12$~kpc where they find 2--4 stream stars with
$v_{\rm hel}\sim-500$~km~s$^{-1}$ and a group of 7~stars at $v_{\rm
hel}\sim-400$~km~s$^{-1}$.  These data are shown in the top panel of
Figure~\ref{fig:ibata}.

\begin{figure}
\begin{center}
\leavevmode
\includegraphics[width=9cm]{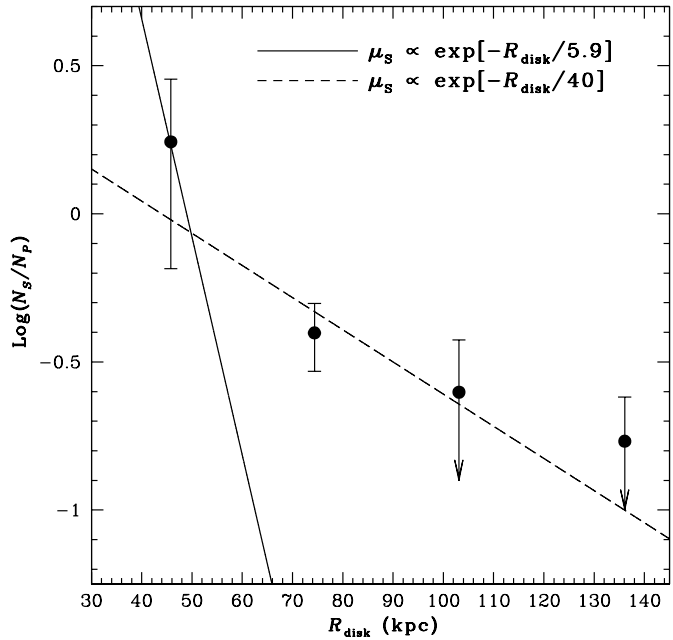}
\end{center}
\caption{Ratio of stars in the secondary velocity peak to those in the
primary velocity peak plotted as a function of radius in the plane
of M31's disk (see \S\,\ref{interpdisk}).  The data points are found to be
inconsistent with a smooth disk following a $\mu_S\propto{\rm exp}(-R/R_{\rm
exp})$ surface brightness law with $R_{\rm exp}=5.9$~kpc (solid line).  This
profile, as well as the profile with $R_{\rm exp}=40$~kpc, have both been
arbitrarily scaled in the vertical direction.  The outer two data points are 
shown as upper limits, with an arrow signifying the size of the random 
error bar.} 
\label{fig:ratio}
\end{figure}

The histograms in Figure~\ref{fig:ibata} show mild evidence of a trend
similar to what we would expect if the secondary peak consists of M31 disk
stars.  The ratio of the number of stars in the secondary velocity peak
(between the dashed lines) to the primary peak appears to decrease as a
function of increasing radius.  We have defined the secondary peak here based 
on the Ibata et~al.\ (2004) F8 field, which shows the largest contrast between 
secondary and primary populations.  To a first approximation, we can quantify
this trend by simply counting the number of stars in each component directly
from Figure~\ref{fig:ibata}.  Over the projected radial distance of the
stream in the four~pointings shown in Figure~\ref{fig:ibata} (12--33~kpc),
the star-count map of M31 (Figure~1) does not show any obvious changes in the
surface density of the giant southern stream.  If we assume that the surface
density of the giant southern stream (primary peak) is constant, then the
ratio of secondary peak to primary peak stars effectively gives us the radial
surface brightness profile of the secondary component.  Since the apparent
brightness and metallicity ranges of stars in both peaks is virtually
identical, taking the ratio also eliminates any possible biases (e.g., there
may be variations in the degree of completeness from one study to another,
but this should affect both primary and secondary peak populations equally).
If the outer disk is responsible for the secondary peak, then the radial
dependence of this surface brightness should follow the measured exponential
profile of M31's disk.  We compute $R_{\rm disk}$ for each of the fields in
Figure~\ref{fig:ibata} as discussed earlier for H13s.  A radial density
distribution based on these numbers is shown in Figure~\ref{fig:ratio}.  Also
shown is an exponential disk with $\mu_S\propto{\rm exp}(-R_{\rm
disk}/5.9$~kpc) (solid).  This exponential disk profile for this scale length is 
in gross disagreement with the radial secondary peak data, confirming that this 
population does not look like the smooth disk of M31.  For comparison, the dashed 
line illustrates that in order to fit these data with such a model, the scale 
length of the disk would have to be very large: $R_{\rm exp}\approx40$~kpc.  Adopting a
less extreme inclination for the disk --- say a flattening of
$(b/a)\equiv1-\epsilon=0.31$ (Worthey et~al.\ 2005) --- does not resolve the
mismatch.  Since there is no convincing evidence for a secondary population in 
the bottom two panels of Figure~\ref{fig:ibata}, we have plotted these data points 
as upper limits in Figure~\ref{fig:ratio}.  The size of the arrow indicates the 
random error associated with these points.  We note however that if the surface 
density of the giant southern stream (primary peak) decreases with increasing 
distance from the plane of M31, then our outer data points have been artificially 
``boosted'' and a model with a somewhat smaller scale length will fit the 
secondary peak data.  We also note that a very small number of smooth spheroid 
stars may exist in each of the primary and secondary velocity peaks and 
have not been removed.  This contribution is known to be very small and does not 
affect the results.

Taken together, both the surface brightness and radial density distribution
of the secondary peak in H13s argue against it consisting of M31 smooth disk
stars.  It is however possible that this cold population represents some kind
of disk feature, possibly tidally ripped from M31's smooth disk during
a satellite accretion event.  Such a warp is known to exist along the major 
axis of M31 (e.g., the Northern Spur) but has not been reported on the minor 
axis.

\section{Future Studies} \label{future}

Despite the many studies that have looked at Andromeda's giant southern stream,
several important issues remain unresolved including the exact orbit of the
stream and the nature of the progenitor system that created it.  The secondary
peak detected in this study adds to the observational database and will surely
invite different groups to construct new models to explain this feature.
Future Keck/DEIMOS 
observations should target spectroscopy fields in the inner regions of the giant southern 
stream to improve the statistics currently constrained solely by the Ibata et~al.\ (2004) 
detection (a few stars in the stream at $R\sim12$~kpc).  The multi-object 
spectroscopic method employed in this study will be very efficient in the inner 
regions of the stream and should detect $\sim100$ bonafide stream stars per DEIMOS 
mask.  It would also be interesting to conduct observations in a field directly
to the north-east of our current pointing, stepping by $\approx1$~degree
parallel to the major axis.  Observing a field at the same radial distance, 
but off the stream, should confirm the presence of the secondary peak found in
this study if it is related to the disk of M31. 

Recent ultra-deep {\it HST\/} observations have now been taken in overlapping
fields to this study (T.\ Brown 2005, private communication).  These
observations, in H13s, H11, and H13d will complement our study by providing
additional constraints.  The mean ages of stars in these fields, and the spread
in ages, will be directly measured from fitting the morphology of the
main-sequence turnoff in those data with isochrones and by comparing the
numbers of stars in various evolutionary phases with numerical simulations.  A
relative comparison of the CMDs of the H13s and H13d fields could potentially
help resolve the stream from the outer disk in these inner regions.

Wide-field CFHT/MegaCam images (1~degree$^2$) have also been obtained directly
covering our H13s and H11 fields.  These images will provide much deeper star counts than in the 
Ibata et~al.\ (2005) map and will allow us to search for lower surface brightness 
features in these fields.  The MegaCam images will directly provide detailed starcounts as 
a function of radius over the inner regions of the giant stream.  These will be very important 
in the face of our discovery of a secondary population in our H13s field.  The new star-count map 
will also better define the width and northern sharp edge of the giant stream, both of which are 
important inputs into constraining models for the stream's orbit.  These observations are 
currently being reduced and will be published in a forthcoming paper.


\section{Conclusion} \label{conclusion}

We present Keck/DEIMOS observations of red giant stars in three~fields in the
Andromeda spiral galaxy: a field on the giant southern stream, a minor-axis 
spheroid field, and a major-axis disk field.  In this paper, we discuss
the kinematics and chemical composition of stars in the stream field at a
projected distance of $R\sim21$~kpc from M31's center.  The disk and spheroid
fields will be addressed in future papers.  

We isolate RGB stars in M31 by removing foreground Galactic dwarf star
contaminants through probability distribution functions calculated from a
training set of known RGB and dwarf stars.  These functions combine
four~individual parameters --- (1)~radial velocity, (2)~Na\,{\sc i}
equivalent width (surface gravity sensitive), (3)~position in the CMD, and
(4)~comparison between photometric and spectroscopic [Fe/H] estimates --- to
compute a likelihood that any given object is an M31 RGB star or a Milky Way
dwarf star.  The M31 RGB sample shows the cleanest detection to date of the 
inner part of the giant southern stream: a kinematically-cold population
($\sigma(v_1)=16$~km~s$^{-1}$) of 50~stars moving with a mean heliocentric
velocity of $\bar{v_1}=-513$~km~s$^{-1}$.  These stars are found to be
metal-rich, with a mean metallicity $\rm\langle[Fe/H]\rangle=-0.47$, and a
small metallicity dispersion.  We also find clear evidence for a second
population of stars that is also moving with a relatively large negative
radial velocity with respect to M31's systemic velocity:
$\bar{v_2}=-417$~km~s$^{-1}$ that is comparably cold:
$\sigma(v_2)=16$~km~s$^{-1}$).  This second cold population has never been
reported before.  A close look at previous spectroscopic observations reveals
evidence that this second group of stars is not associated with the smooth
disk of M31, but rather may represent a new satellite debris trail, a wrapped around 
component of the giant southern stream, or a tidally-disrupted population of disk 
stars.  Future observations of the inner regions of the stream will help constrain 
models for its orbit and put better limits on which, if any, of M31's satellites 
may be the progenitor.


\acknowledgments

We wish to thank Peter Stetson and Jim Hesser for help in acquiring the
CFHT/MegaCam imaging fields for this project.  We are also grateful to Peter
Stetson and James Clem for providing programs and for many useful discussions
regarding the astrometry of the CFHT images.  We wish to thank Carynn Luine
for help with the verification of the radial velocity measurements and Leo
Girardi for providing us with an extensive grid of theoretical stellar
isochrones.  We also acknowledge Tom Brown for making available unpublished 
results based on Cycle 13 {\sl HST} observations.  J.S.K. is supported 
by NASA through Hubble Fellowship grant HF-01185.01-A, awarded by the Space
Telescope Science Institute, which is operated by the Association of 
Universities for Research in Astronomy, Incorporated, under NASA 
contract NAS5-26555.  This project was also supported by NSF grant 
AST-0307966 and NASA/STScI grant GO-10265.02 (J.S.K., P.G., and 
K.M.G.), an NSF Graduate Fellowship (K.M.G.), NSF grant AST-0307931 (R.M.R. and 
D.B.R.), and NSF grants AST-0307842 and AST-0307851, NASA/JPL contract 1228235, 
the David and Lucile Packard Foundation, and The F.~H.~Levinson Fund of the 
Peninsula Community Foundation (S.R.M.).

\clearpage

\notetoeditor{We would prefer that Figure 1 be printed on at least half of one column in the 
two column format.  However, Figures 3 and 4 look nice if they occupy half of the page, going 
across both columns}

\clearpage

\clearpage

\begin{deluxetable}{cccccccccc}
\rotate
\tabletypesize{\scriptsize} \tablecaption{Keck/DEIMOS Observations of
M31\tablenotemark{1}
\label{table1}}
\tablewidth{0pt}
\tablehead{\colhead{Field Location} & \colhead{Field ID} & \colhead{Mask ID}
& \colhead{$\alpha$ (J2000)} & \colhead{$\delta$ (J2000)} & \colhead {$R_{\rm proj}$} & 
\colhead{Pos.\ Angle} & \colhead{Exp.\ Time}\tablenotemark{2} & \colhead{No.\
Targets\tablenotemark{2}} & \colhead{Airmass} \\ \colhead{} & \colhead{} & \colhead{} & \colhead{(h:m:s)} &
\colhead{($^\circ$:$'$:$''$)} & \colhead{(kpc)} & \colhead{($^\circ$)} & \colhead{(s)} & \colhead{} & \colhead{}}
\startdata

Giant Southern Stream & H13s & f2\_1 & 00:44:14.76 & +39:44:18.2 & 21 & +21.0 &
3600 & 133 & 1.84 \\
& & f2\_2 & 00:44:14.76 & +39:44:18.2 & 21 & $-$21.0 & 3600 & 137 & 1.33 \\

\\
South-East Minor Axis & H11 & f1\_1 & 00:46:21.02 & +40:41:31.3 & 12 & +21.0 &
3600 & 136 & 1.15 \\
(Spheroid) & & f1\_2 & 00:46:21.02 & +40:41:31.3 & 12 & $-$21.0 & 3600 & 135 &
1.07 \\

\\
North-East Major Axis & H13d & f3\_1 & 00:49:04.80 & +42:45:22.6 & 26 & +27.8 &
3600 & 145 & 1.10 \\
(Disk) & & f3\_2 & 00:49:04.80 & +42:45:22.6 & 26 & +27.8 & 3600 & 144 & 1.32 \\
\enddata

\tablenotetext{1}{All observations were obtained on 2004 September~20 (UT).} 
\tablenotetext{2}{2 targets were observed twice on each of H13s and H11, and 53 
targets were observed twice on H13d; see \S\S\,\ref{masks} \& \ref{velocityquality}.} 



\end{deluxetable}

\clearpage

\end{document}